\documentclass[sigconf]{acmart}
\AtBeginDocument{%
 }

\copyrightyear{2023}
\acmYear{2023}
\setcopyright{rightsretained}
\acmConference[GLSVLSI '23]{Proceedings of the Great Lakes Symposium on VLSI 2023}{June 5--7, 2023}{Knoxville, TN, USA}
\acmBooktitle{Proceedings of the Great Lakes Symposium on VLSI 2023 (GLSVLSI '23), June 5--7, 2023, Knoxville, TN, USA}
\acmDOI{10.1145/3583781.3590266}
\acmISBN{979-8-4007-0125-2/23/06}



\usepackage{amsmath,amsfonts}
\usepackage{mathtools}
\usepackage{algorithmic}
\usepackage{graphicx}
\usepackage{caption}
\usepackage{textcomp}
\usepackage{xcolor}
\usepackage{comment}
\usepackage{tabularx,ragged2e,booktabs}
\usepackage{makecell}
\usepackage{multirow}
\usepackage[most]{tcolorbox}
\usepackage[justification=centering]{caption}
\usepackage[utf8]{inputenc}
\usepackage{longtable}
\usepackage{caption}
\usepackage{lipsum}% http://ctan.org/pkg/lipsum
\usepackage{cleveref}
\begin{document}

\title{Island-based Random Dynamic Voltage Scaling vs ML-Enhanced Power Side-Channel Attacks}

\author{Dake Chen}
\affiliation{%
  \institution{University of Southern California}
  \city{Los Angeles}
  \state{California}
  \country{USA}}
\email{dakechen@usc.edu}

\author{Christine Goins}
\affiliation{%
  \institution{Niobium Microsystems, Inc.}
  \city{Dayton}
  \state{Ohio}
  \country{USA}}
\email{christine@niobiummicrosystems.com}

% Max is also USC alumni, so I used USC as his affiliation
\author{Maxwell Waugaman}
\affiliation{%
  \institution{Independent Researcher}
%  \city{Los Angeles}
  \state{Washington}
  \country{USA}}
\email{wellwaugaman@gmail.com}

\author{Georgios D. Dimou}
\affiliation{%
  \institution{Niobium Microsystems, Inc.}
  \city{Dayton}
  \state{Ohio}
  \country{USA}}
\email{georgios@niobiummicrosystems.com}

\author{Peter A. Beerel}
\affiliation{%
  \institution{University of Southern California}
  \city{Los Angeles}
  \state{California}
  \country{USA}}
\email{pabeerel@usc.edu}

\renewcommand{\shortauthors}{Dake Chen, Christine Goins, Maxwell Waugaman, Georgios D. Dimou, \& Peter A. Beerel}

\begin{abstract}
  A clear and well-documented \LaTeX\ document is presented as an
  article formatted for publication by ACM in a conference proceedings
  or journal publication. Based on the ``acmart'' document class, this
  article presents and explains many of the common variations, as well
  as many of the formatting elements an author may use in the
  preparation of the documentation of their work.
\end{abstract}




\begin{abstract}
In this paper, we describe and analyze an island-based random dynamic voltage scaling (iRDVS) approach to thwart power side-channel attacks. We first analyze the impact of the number of independent voltage islands on the resulting signal-to-noise  ratio and trace misalignment. As part of our analysis of misalignment, we propose a novel unsupervised machine learning (ML) based attack that is effective on systems with three or fewer independent voltages. Our results show that iRDVS with four voltage islands, however, cannot be broken with 200k encryption traces, suggesting that iRDVS can be effective. We finish the talk by describing an iRDVS test chip in a 12nm FinFet process that incorporates three variants of an AES-256 accelerator, all originating from the same RTL. This included a synchronous core, an asynchronous core with no protection, and a core employing the iRDVS technique using asynchronous logic. Lab measurements from the chips indicated that both unprotected variants failed the test vector leakage assessment (TVLA) security metric test, while the iRDVS was proven secure in a variety of configurations.
\end{abstract}

\maketitle

\section{Introduction}
\label{sec:intro}

With billions of computing devices being deployed in the Internet of Things, autonomous vehicles and other pervasive applications,
%As our world embraces the internet of things and autonomous vehicles, 
ensuring our integrated circuits are secure has become as important as improving performance, power, and area. In particular, power side-channel attacks have become an increasing source of concern~\cite{kocher1999differential}. 

Various countermeasures against power side-channel attacks have been proposed, including masking to make power signatures independent from cryptographic keys \cite{detrano_exploiting_2015}, current flattening using sophisticated power delivery mechanisms \cite{kar_invited_2017}, and 
increasing the difficulty of  trace alignment, a critical step in power attacks~\cite{lu_fpga_2008, jayasinghe2019scrip}.
%specialized logic families, including dual-rail \cite{tiri2004logic} and power-balanced logic \cite{BH2015}.
%proposed power balanced logic against static power attack .%All ares relatively expensive to implement, either in design time or area/power consumption.
%Other proposed countermeasures try to protect systems by increasing the difficulty of  trace alignment, a critical step in power attacks~\cite{lu_fpga_2008, jayasinghe2019scrip}. 
%These techniques include inserting random delays~\cite{lu_fpga_2008}, using random clocks~\cite{guneysu_generic_2011, jayasinghe2019scrip}, and using globally asynchronous locally synchronous (GALS) design \cite{soares_robust_2011}.
Alignment techniques have co-evolved with these countermeasures. 
Powerful approaches for aligning traces include ``time warping'' techniques, such as elastic alignment~\cite{woudenberg_improving_2011} and the rapid alignment method \cite{bayrak_eda-friendly_2013}. 
%Sliding window approaches analyze windows of a signal that are likely to include the secret information  \cite{lellis_energy-based_2017}.
%Other methods extract the peak values for each cycle and use them to reconstruct and align the traces for power attacks \cite{tian_power_2012}.
%More advanced approaches for aligning traces include ``time warping'' techniques, such as elastic alignment~\cite{woudenberg_improving_2011} and the rapid alignment method \cite{bayrak_eda-friendly_2013}. 
The countermeasure closest to that proposed in this paper is \emph{dynamic voltage scaling (DVS)}. DVS affects both the timing and amplitude of power traces \cite{baddam_evaluation_2007,yu_false_2017,singh_improved_2017}. 
However, most DVS approaches have been restricted to using a single voltage for the entire design and are vulnerable to the estimation of the random voltage \cite{baddam_evaluation_2007}. Machine learning techniques have been extensively used for attacking hardware security systems~\cite{10129346}.
In particular, with the help of the unsupervised machine learning algorithm, we show that DVS can be effectively attacked by grouping the power traces into clusters with similar supply voltages and attacking one cluster. 

This paper proposes and analyzes an \emph{island-based random dynamic voltage scaling (iRDVS)} approach that uses multiple independent random voltages that are more difficult to estimate. We first analyze the \emph{signal-to-noise ratio (SNR)} of iRDVS, then analyze the resistance of this technique to alignment. We evaluate both as a function of the number of independent voltages. Together, we argue, these analyses suggest that a design with a small number of independent voltages achieves high security.

As part of our alignment analysis, we propose a novelly applied unsupervised ML algorithm to cluster iRDVS traces and enable more effective power attacks. 
%Clustering algorithms have been used to create profiles of given hardware using traces with known keys, as well as to identify regions of interest in encryption algorithms that use exponentiation \cite{heyszl2013clustering}. %\cite{heyszl2013clustering, perin2014attacking}.
Our approach, %by contrast, 
uses clustering to classify power traces from different unknown iRDVS voltages into groups of similar voltages. Our experimental results show that this clustering-based attack is able to uncover keys in systems protected by one, two and three dynamic voltage islands but has limited benefit when applied to iRDVS schemes with four or more islands.

The rest of the paper is organized as follows. Section~\ref{sec:backg} describes the techniques and metrics used in this paper. 
\Cref{sec:irdvs} proposes the iRDVS design. \Cref{sec:snra} and \Cref{sec:temp} analyze the SNR and misalignment characteristics of our approach as a function of the number of islands, and \Cref{sec:clusteringattack} focuses on our proposed ML attack. \Cref{sec:experiment} and \Cref{sec:chip} describes the details of our simulation experiments and measurement results. Finally, we provide a summary of this work with our plans for future work in \Cref{sec:concl}.

\section{Background}
\label{sec:backg}
% will brief the technique, tool, algorithms we leveraged in this paper

This section summarizes correlation-based power analysis, provides details on the elastic alignment technique used to test our approach, and introduces three common metrics for quantifying countermeasure effectiveness.

\subsection{Correlation-Based Power Analysis}
Power analysis attacks take advantage of the dependence of a circuit’s power consumption on the data it processes. A common method of disclosing this correlation employs a differential technique introduced by %\citeauthor{kocher1999differential} 
Kocher et al. \cite{kocher1999differential} called \emph{differential power analysis (DPA)} that recovers keys bit-by-bit. Another powerful technique requiring less knowledge of the algorithm implementation, introduced by %\citeauthor{brier2004correlation}
Brier et al., is \emph{correlation-based power analysis (CPA)} \cite{brier2004correlation}. %\citeauthor{brier2004correlation}
Brier et al. demonstrated that all countermeasures against CPA provide similar defensive effectiveness against DPA. Moreover, CPA is capable of attacking several bits at a time instead of only a single bit.
Because of this advantage, we applied CPA in our experiments.

\subsection{Elastic Alignment}

%\citeauthor{woudenberg_improving_2011}
Woudenberg et al.~\cite{woudenberg_improving_2011} propose a powerful alignment algorithm, \emph{elastic alignment}, to preprocess traces corrupted by random delay insertion or an unstable clock.
The two step procedure aligns recorded traces to a single reference. First, it leverages a traditional algorithm, \emph{dynamic time warping}, to find a {\em warp path} that maps the time steps of each sample trace to those of a reference trace. 
To do this, the algorithm computes the Euclidean difference between each target trace $t$ and a reference trace $r$,  
captured in a 2-D cost matrix
of size $P \times Q$, where $P$ and $Q$ are the lengths of traces $t$ and $r$, respectively.
It then applies dynamic programming to identify the minimum-cost path through the matrix between points $(0,0)$ and $(P,Q)$.
%to be the warp path.
This path defines the correspondence between traces $r$ and $t$.
Secondly, guided by this path, elastic alignment averages across samples when multiple samples of $t$ map to one time step and duplicates samples of $t$ when one sample of $t$ maps to multiple time steps. We reproduce this approach and show it can effectively align the traces corrupted by the frequency-scaled technique.

\begin{comment}

\begin{figure}[hbt!]
    \centering
    \includegraphics[width=9cm]{Fig1.png}
    \caption{Elastic alignment applied to a trace misaligned with dynamic frequency scaling}
    %\vspace{-3mm}
    \label{fig:warpingdfs}
\end{figure}
% show the function of warping technique first
\Cref{fig:warpingdfs} illustrates the alignment process.
The first trace shown in \Cref{fig:warpingdfs} is a reference trace from \cite{luo2014side}.
The second trace is a frequency-scaled version of the first, and the third trace is the result of applying the elastic alignment algorithm to the second trace, illustrating a successful alignment. 
\end{comment}
%The associated warped path is shown in Fig. \Cref{fig:warpingdfs}(b) in which the time samples of the reference trace $r$ are on the x-axis and the time samples of the misaligned trace $t$ on the y-axis. The horizontal segments of the warp trace indicates where the warping algorithm maps a single sample of the faster trace $t$ to multiple samples of the slower reference trace $r$.

\vspace{-6pt}
\subsection{Metrics for Countermeasure Effectiveness}

There are three common metrics for measuring side-channel countermeasure effectiveness. The first is \emph{Minimum Traces to Disclosure (MTD)}, which is the number of encryption/decryption traces required to disclose all of the secret information. A higher MTD indicates a more secure countermeasure. This metric requires that all bytes of the secret are guessed correctly.
\emph{Partial Guessing Entropy (PGE)} \cite{oflynn_side_2014} can be a more practical evaluation metric than MTD because it does not require a correctly-guessed secret. PGE is computed from the ranking of possible values of the subkey bytes in descending order of correlation as estimated by the Pearson correlation coefficient. PGE is the rank of the correct subkey, where a PGE of 0 denotes that the subkey was correctly guessed.
A large PGE indicates a low correlation of the correct subkey and consequently a system robust to attacks.  
The \emph{Test Vector Leakage Analysis (TVLA)} is also commonly used to evaluate side-channel leakage~\cite{Becker2013TestVL}. 
The test conducts two experiments with fixed and random plaintexts and generates a large number of traces respectively, and the power samples from two groups are used for calculating a t-score at each time step. The t-score is a statistical measure of how different traces from the fixed and random encryptions are from one another. A higher t-score indicates that the difference between the fixed and random traces is less likely to have occurred by chance, suggesting
the device exhibits higher leakage that would make a power analysis attack more likely to succeed.  
%T-scores are calculated point-wise, so each point in time in which a power sample is taken has its own t-score. 

\section{Island-based Random DVS}
\label{sec:irdvs}
%\vspace{-2mm}

Traditional DVS countermeasures can be attacked if the random dynamic voltage is uncovered \cite{baddam_evaluation_2007}. Attackers can scale measured power traces in time and amplitude to match a reference trace, which renders DVS designs vulnerable.

\begin{figure}[hbt!]
    \centering
    \includegraphics[width=5.5cm]{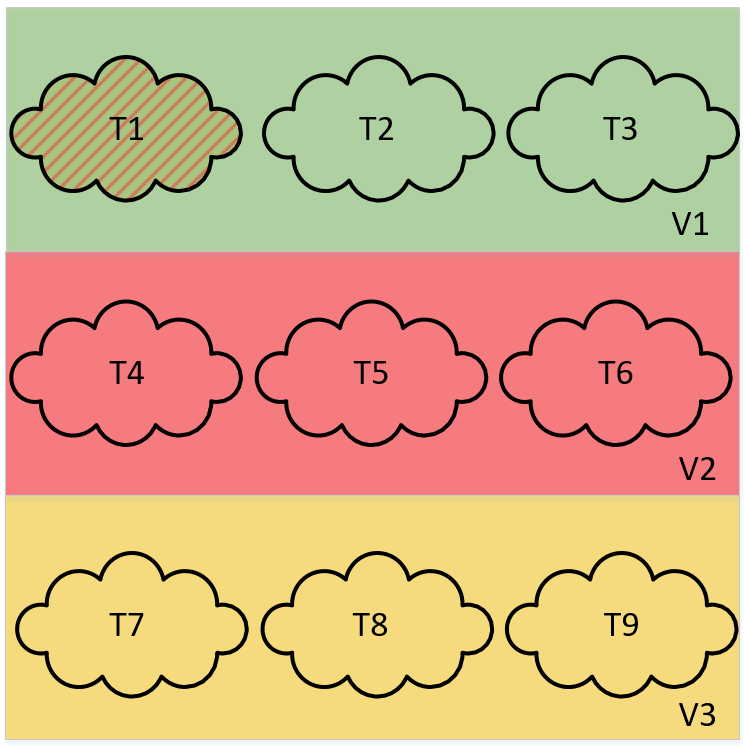}
    \caption{Illustration of a typical iRDVS structure with $n = 9$ islands and $m = 3$ independent voltages. Each independent voltage domain has a different color and each cloud represents a group of logic; the shaded logic is under attack.}
    %\vspace{-3mm}
    \label{fig:irdvs_structure}
\end{figure}

To circumvent the weaknesses of single-island DVS, this paper proposes using several independent voltages in an \emph{island-based random DVS (iRDVS)} framework, illustrated in \Cref{fig:irdvs_structure}.
iRDVS makes side-channel attacks more difficult because attackers must differentiate between multiple simultaneous random dynamic voltages.

One practical means of implementing this iRDVS framework for a pipelined design is to partition each combinational stage into multiple islands with independent voltages. The voltages can be randomly adjusted with the constraint that the delay of each pipeline stage is roughly the same, thereby maximizing overlapping computation and minimizing the chance of introducing timing side channels. 
Multiple islands can share one voltage supply to support scaling this approach to large circuits with many islands. The islands need not all be the same size but can be adjusted based on 
both logical and physical constraints and communicate using asynchronous channels that implement 
the flow-control necessary to cope with different stage delays \cite{beerel_book}
We assume each island will have an on-chip DC/DC converter whose control will leverage the entropy from an off-the-shelf true random number generator (TRNG). This is similar to the random voltage generation proposed for random DVS \cite{liu2012true, yu_exploiting_2018, yang_power_2005}. As was implemented in \cite{liu2012true}, we assume the TRNGs will be on-die and thus not directly accessible to power attacks.
Determining the optimal number and configuration of independent voltages that not only thwarts voltage prediction but also retains the statistical merits of DVS is one of the key research objectives we explore here.

\begin{comment}
In the iRDVS design, the power consumption of one island helps obfuscate the power of other islands. %, the overlapped power signature renders the traces of iRDVS confusing. 
Compared to a single DVS, the randomness of the power signatures in iRDVS can be intensified but the relative variance of the secret island to the noise, a key factor in the viability of power attacks, is less obvious (see \Cref{sec:snra}).
For the implementation of encryption algorithms, only a portion of the power, referred to as the region of interest, stems from the secret key; this portion is referred to as a \emph{region of interest}. Therefore, attackers usually first identify a region of interest that contains the secret and extract it from every trace \cite{bayrak_eda-friendly_2013}. 
%This pre-preprocess greatly reduce the timing and space complexity of power attacks. For this purpose, attackers exploit  pattern matching algorithms.
%to pinpoint it. %These algorithms mainly apply different numerical difference or distance equations to measure how similar two portions are.f
The iRDVS framework contains randomness in its traces, which hinders identification of regions of interest.
%functions ineffective, thereby thwart this kind of pre-process. 
\end{comment}

\section{SNR Analysis}
\label{sec:snra}

%The randomness of power amplitude injected by iRDVS is a fortress resisting power attacks, the optimal number of islands become a critical point which decides the robustness of an iRDVS system.
The \emph{signal-to-noise ratio (SNR)} is typically used to quantify how well the secret portion of the computation is hidden within the overall power consumption \cite{mangard2004hardware}. 
The SNR is defined as $SNR=\frac{Var(AP)}{Var(N)}$
\begin{comment}
\begin{align}
    SNR=\frac{Var(AP)}{Var(N)} \
\end{align}
\end{comment}
where $AP$ denotes the power consumption associated with the intermediate value that carries secret information and $N$ consists of the power consumption of uncorrelated computations and electronic noise. 
In this section, we examine the SNR of various island configurations to analyze their effectiveness.
To simplify our analysis, we assume the traces are perfectly aligned; we analyze the misalignment benefit associated with iRDVS in the next section. 

The correlation between the hypothetical intermediate value and power traces can be derived in terms of SNR: 
\begin{align}
    \rho=\frac{\rho_{ap}}{\sqrt{1+\frac{1}{SNR}}} \
\end{align}
where $\rho_{ap}$ denotes the correlation between the power consumption of the attacked part and the hypothetical intermediate value \cite{mangard2004hardware}. 
This equation shows that a lower SNR leads to a lower correlation, which indicates higher robustness.

Let $T_i$ be a power trace for island $i$ normalized by the voltage of that island. Let $v_i$ denote the independent random dynamic supply voltage for island $i$. Because the switching power is proportional to $v^\alpha$, where $\alpha\approx{}2$, and most instantaneous power consumption is from switching power, the DVS power traces are proportional to $v^\alpha T$.
Let $n$ denote the number of independent islands and $m$ represent the number of independent voltages used. We present three different cases for comparison: first, the $m=n$ independent DVS case, which means we assign a different random voltage to each island; then, the cases with two ($m=2$) and one ($m=1$) independent voltages.

Without loss of generality, assume the first island is attacked, so the power consumption of the other $n-1$ islands is switching noise.
The SNR for $m=n$ iRDVS islands ($v_1$, $v_1$, ..., $v_n$) can be represented as follows. 
Let $\sigma$ and $\mu$ denote the standard deviation and mean of their associated variables, respectively.
Since $v_i$ and $T_i$ are independent of each other, we can expand the variance for both denominator and numerator.

{%\small
\begin{align}
 SNR_{m=n}&=\frac{Var(v_1^\alpha T_1)}{Var(\sum_{i=2}^{n} v_i^\alpha T_i)}
 \\&=\frac{\sigma^2_{v_1^\alpha} \sigma^2_{T_1}+\sigma^2_{v_1^\alpha} \mu^2_{T_1}+\mu^2_{v_1^\alpha} \sigma^2_{T_1}}{\sum_{i=2}^{n}(\sigma^2_{v_i^\alpha} \sigma^2_{T_i}+\sigma^2_{v_i^\alpha} \mu^2_{T_i}+\mu^2_{v_i^\alpha} \sigma^2_{T_i})}
\end{align}
}

Considering the special case that variances and means of the island power consumption and supply voltages are the same, denoted $\sigma_T^2$, $\mu_T$, $\sigma^2_{v^\alpha}$ and $\mu_{v^\alpha}$, we obtain:
{%\small
\begin{equation}
 SNR_{m=n}=\frac{\sigma^2_{v^\alpha} \sigma^2_{T}+\sigma^2_{v^\alpha} \mu^2_{T}+\mu^2_{v^\alpha}\sigma^2_{T}}{{ (n-1)}\sigma^2_{v^\alpha} \mu^2_{T}+(n-1)(\mu^2_{v^\alpha} \sigma^2_{T}+\sigma^2_{v^\alpha} \sigma^2_{T})} %\label{eq:first}
\end{equation}
}

Similarly, we can derive the SNR for the cases where $m$ is equal to two ($v_1$, $v_2$) and one ($v$) independent voltages, the latter modeling the conventional DVS approach. 

{\small
\begin{align}
 SNR_{m=2}&=\frac{Var(v_1^\alpha T_1)}{Var(v_1^\alpha \sum_{i=2}^{\frac{n}{2}}  T_i+v_2^\alpha \sum_{i=\frac{n}{2}+1}^{n}  T_i)} \nonumber
 \\&=\frac{\sigma^2_{v^\alpha} \sigma^2_{T}+\sigma^2_{v^\alpha} \mu^2_{T}+\mu^2_{v^\alpha}\sigma^2_{T}}{{ [(\frac{n}{2}-1)^2+(\frac{n}{2})^2]}\sigma^2_{v^\alpha} \mu^2_{T}+(n-1)(\sigma^2_{v^\alpha} \sigma^2_{T}+\mu^2_{v^\alpha} \sigma^2_{T})}
 %\label{eq:second}
\end{align}
}

%\small
\begin{align}
 SNR_{m=1}&=\frac{Var(v_1^\alpha T_1)}{Var(v_1^\alpha \sum_{i=2}^{n}  T_i)} \nonumber
 \\&=\frac{\sigma^2_{v^\alpha} \sigma^2_{T}+\sigma^2_{v^\alpha} \mu^2_{T}+\mu^2_{v^\alpha}\sigma^2_{T}}{{ (n-1)^2}\sigma^2_{v^\alpha} \mu^2_{T}+(n-1)(\sigma^2_{v^\alpha} \sigma^2_{T}+\mu^2_{v^\alpha} \sigma^2_{T})} %\label{eq:third}
\end{align}

%For the denominator of both cases, we factored the supply voltage out of the summation of the switching activities of the same voltage. 
Due to the algebraic property that for $a\geq1$ and $b\geq1$, $(a+b)^2\geq a^2+b^2 \geq a+b$, it can easily be shown that $SNR_{m=n} \geq SNR_{m=2} \geq SNR_{m=1}$. %if we assume the variances and  means  of  the  power  consumption and supply voltages are the same.
This indicates that, 
somewhat counter-intuitively, without considering the misalignment and temporal advantage, a \textit{lower} number of DVS islands results in lower SNR, thereby lower correlation and higher robustness. 

We can explain this trend more generally from the perspective of covariance:
\begin{align}
    Var(v^\alpha_i T_i+v^\alpha_j T_j)
    =&Var(v^\alpha_i T_i)+Var(v^\alpha_j T_j)+\\&2Cov(v^\alpha_i T_i,v^\alpha_j T_j)
    \nonumber
\end{align}

The above equation is the general formula for computing the variance of two islands. If the two islands have the same supply
voltage, i.e., $v_i=v_j$, the two quantities $v^\alpha_i T_i$ and $v^\alpha_j T_j$ are correlated, therefore the covariance term $Cov(v^\alpha_i T_i,v^\alpha_j T_j)$ is greater than zero. When the two islands have independent random voltages with possibly different means and variances, i.e., $v_i\neq v_j$ where $\sigma_T^2$, $\mu_T$, $\sigma^2_{v^\alpha}$ and $\mu_{v^\alpha}$ are not equal, the two quantities $v^\alpha_i T_i$ and $v^\alpha_i T_i$ are also independent and $Cov(v^\alpha_i T_i,v^\alpha_j T_j)=0$. %Considering the covariance, even if the assumption does not hold, i.e., $T_i$ and $T_j$ are not independent, the covariance of case would be $v_i\neq v_j$ still less than the case $v_i=v_j$, so the conclusion  $SNR_{m=n} \geq SNR_{m=2} \geq SNR_{m=1}$ holds.

This reduction in variance caused by an increasing number of supply voltages can be generalized to more islands. %($n$) case.
For the case of $m=2$, the $\frac{n}{2}$ noise islands supplied by $v_2$ are correlated with one another, increasing the covariance terms in the denominator and decreasing the SNR. However, if we keep $n$ constant and increase the number of voltage supplies $m$, the correlation between islands reduces (increasing SNR) because fewer islands are correlated with each other.
When $m=n$, each island is powered by an independent supply so there is no covariance among noise islands. Therefore, the variance of the noise is minimized and the SNR is maximized for this $n$. 

We experimentally verified this trend by performing CPA on a simplified model of AES that simulated the Sbox operations in the first round of AES.
This model, given a plaintext, computes the 16 Sbox output values for the first round of AES and generates a power pulse with a peak amplitude corresponding to the sum of their Hamming weights and a fixed width.
%This pulse was scaled in time according to the Sakurai-Newton delay model with $\alpha = 2$ and in amplitude according to the squared voltage of the island \cite{chandrakasan1992low}.
%\Cref{fig:iRDVS_sims} demonstrates the results of varying the number of independent supplies $m$ in the simplified model of iRDVS.
%The y-axis of each graph specifies the correlation values, where the peaks are located at the correct byte hypotheses for the first byte of the AES key.
%A larger peak at the correct byte hypothesis indicates the design is easier to attack.
The results show that the peak correlations of the single-island and two-island cases are close, and as the number of independent voltages increases, the correlation also increases. However, the correlation for a small number of independent voltages (between 2 and 8) remains well below that observed without dynamic voltage scaling. 

\section{Alignment Analysis}
\label{sec:temp}

iRDVS and DVS introduce temporal advantages for attack resistance in addition to improving SNR by amplitude scaling.
According to the Sakurai-Newton delay model \cite{sakurai1990alpha, chandrakasan1992low}, $ \tau=\frac{C_L V}{k(V-V_T)^\alpha}$, 
\begin{comment}

\begin{align}
\nonumber
    \tau=\frac{C_L V}{k(V-V_T)^\alpha} \ ,
 \nonumber
\end{align}
\end{comment}
the delay of the gates is closely related to the voltage supply.
Assuming each independent voltage can change much faster than the duration of an attack, the power samples of the secret component will be shifted in time as the voltage changes.
This means that the power samples associated with the secret operation, which were expected to be aligned, may be spread over a large range.

The work in \cite{mangard2004hardware} presents a relationship between misalignment and the correlation coefficient:
\begin{align}
    \rho(H,v^\alpha T)=\rho(H, v_s^\alpha T_s)*p*\sqrt{\frac{Var(v_s^\alpha T_s)}{Var(v^\alpha T)}} \
\end{align}
where $H$ represents the Hamming weight or Hamming distance matrix of the hypothetical intermediate value, $v^\alpha T$ denotes the power consumption at a certain time, $v_s^\alpha T_s$ refers to the portion of the power consumption caused by the secret operation, and $p$ denotes the probability that the secret operation is consuming power at the attack time.
Thus, $\rho(H, v_s^\alpha T_s)$ is the correlation for the case where the secret samples are perfectly aligned, whereas $\rho(H,v^\alpha T)$ is the correlation for the full design with misalignment.
Having one or more dynamic voltages would lead to a small $p$ by reducing the probability that secret power samples are self-aligned.
iRDVS and DVS reduce $p$ and decrease the correlation coefficient, making CPA attacks more difficult.

\section{Clustering Attack}
\label{sec:clusteringattack}
% iRDVS resistance to clustering attack
Many alignment techniques, including elastic alignment~\cite{woudenberg_improving_2011}, are based on a notion of a distance between traces.
Power traces from similar operations can be aligned by minimizing the distance between them.
Because voltage scaling increases the distance between operations, and multiple supplies add random noise, these techniques are ineffective when applied to our iRDVS approach, as we will show in \Cref{sec:experiment}.
We propose to strengthen alignment attacks using a novel unsupervised ML-based algorithm that clusters the iRDVS traces into several groups that share similar voltage characteristics. After this clustering, we perform a CPA attack on every cluster and rank the possible subkeys based on their derived correlation coefficients, then average the rank of each possible subkey across all clusters and reorder the subkeys based on the average rank. This new rank order combines the information obtained from all individual attacks on all clusters. We pick the subkey with the lowest average rank to determine MTD, and determine PGE by the final rank of the correct subkey.

We propose using the computationally efficient K-means clustering algorithm \cite{macqueen1967some} to group similar traces. This approach heuristically minimizes the distances among the power values of traces in each cluster. 
A critical parameter for K-means clustering is the number of clusters, which is generally set before the start of the clustering algorithm. We hypothesize that the number of clusters should match the number of different voltage combinations used in the trace set. In this way, each cluster will ideally contain traces from the same specific combination of voltages, ensuring the individual power samples containing the secret key are aligned. 

\begin{comment}
The number of ideal clusters $K$ grows quickly with both the number of independent voltage supplies $m$ and the number of distinct voltages $g$ each voltage supply can support, which means that clusters become smaller (and less likely to reveal the secret key) as the number of voltage supplies grows. 
If each island performs different computations, the number of ideal clusters is $g^m$. If, however, each island performs similar computations, as in the case of our experiments, due to the possible same voltage and similar computations, some combinations of voltages would lead to similar power consumption which should be in the same cluster. Thus the ideal number of clusters $K$ reduces to the number of combinations of $m$ samples from a set of size $g$ with repetition allowed. This can be computed as \cite{rosen2012discrete}
\begin{align}
\nonumber
 K = \binom{m+g-1}{m} \ .
 \nonumber
\end{align}
For example, with three voltage supplies each having five distinct voltages, there are $5^3 = 125$ voltage settings but only  $\binom{3+5-1}{3} = \frac{7!}{3! \; \cdot \; 4!} = 35$ voltage combinations with repetition allowed. 
\end{comment}
Note that having many clusters implies that the average number of traces in each cluster will be $K$ times smaller, reducing the effectiveness of the individual CPA analysis on each cluster. 
This motivates the experimental analysis, presented in the next section, of a range of $K$ values to find the optimal number of clusters. Interesting future work includes using machine learning to guide the choice of $K$.

\begin{figure}
    \centering
    \includegraphics[width=8.8cm]{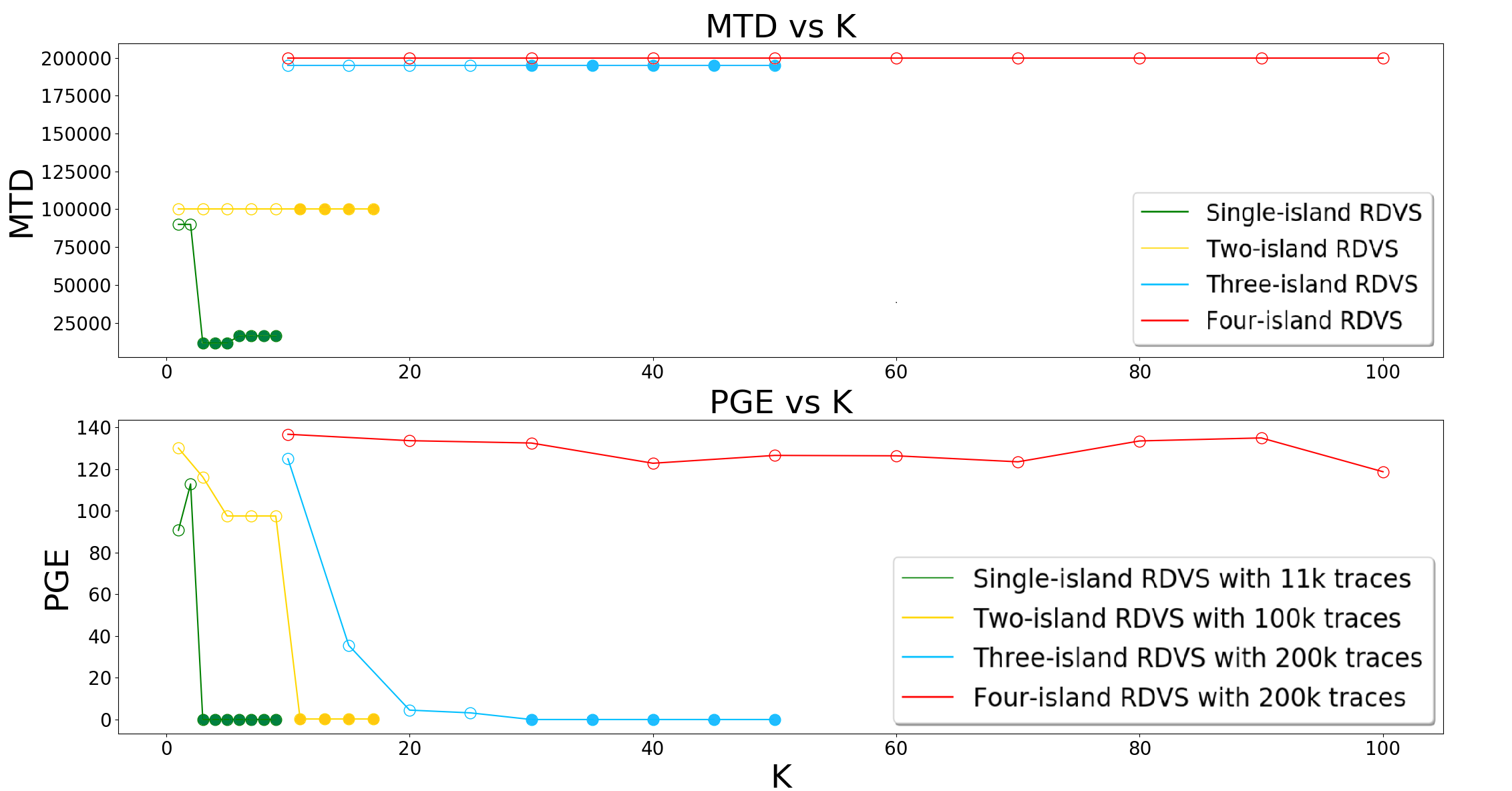}
    \caption{MTD and PGE iRDVS under clustering attack. Empty circles indicate unsuccessful attacks, whereas filled circles indicate successful attacks.}
    %\vspace{-3mm}
    \label{fig:MTD_PGE_Combination}
\end{figure}

\section{Simulation Results and Analysis}
\label{sec:experiment}

This section describes how we evaluated the effectiveness of our iRDVS approach against alignment and CPA attacks.
%technique plus CPA attack and clustering-based attack. 

% show combination and permutation equation: A27

% Based on this results, three or four islands iRDVS have good resistance to clustering attack
\subsection{Trace Generation and Experiment Design} 

We developed an in-house tool in Python to preprocess traces and perform CPA.
Our tool converts power traces from various sources into a standard format, voltage scales the traces, and combines scaled traces to form synthetic iRDVS traces.
It also performs CPA and clustering attacks on both the original and synthetic traces, and generates correlation coefficient, PGE, TVLA, and MTD metrics.
%The tool optionally uses an open-source warping algorithm \cite{FDW} to preprocess the traces.

The original traces used for scaling and combining in these experiments are open-source traces from a combinational 128-bit AES implemented and measured on the Sasebo-GII board by Northeastern University \cite{luo2014side}. 
We make each trace the power consumption of one voltage island.
We use the Sakurai-Newton delay model \cite{sakurai1990alpha} with $\alpha=2$ to expand each sample in the original trace by interpolation,
where $V_{dd}$ for each island is randomly picked from the set $\{0.6, 0.7, 0.8, 0.9, 1\}$.
We then add the scaled traces together to form the synthetic traces of our iRDVS design.
This sum approximates a pipelined implementation of AES where each round operates simultaneously.
To reduce the computation time and increase the probability of disclosure, we also extract the general region of interest from the synthetic traces before running CPA.
Note that the original traces are a set of $100k$, so to generate two island traces, half of the traces are used as signal islands while the other half are used as noise islands, combined into a total of $50k$ traces. If we need more than $50k$ traces, we repeat this process with different scaling and combining of both the signal and noise traces.
This means that the $50k$ signal plaintexts are repeated, but scaled differently and combined with different noise islands. For other multi-island traces, we applied similar methods to generate synthetic traces.

\subsection{Effectiveness of Elastic Alignment}

As described in \Cref{sec:backg}, we tested preprocessing the traces using the open-source Python package {\em fastdtw} \cite{salvador2007toward} to find the warp path, after which we applied elastic alignment based on
%\citeauthor{woudenberg_improving_2011}
Woudenberg et al.'s approach \cite{woudenberg_improving_2011}. 

%\vspace{-2mm}
\begin{figure}[hbt!]
    \centering
    \includegraphics[width=6cm]{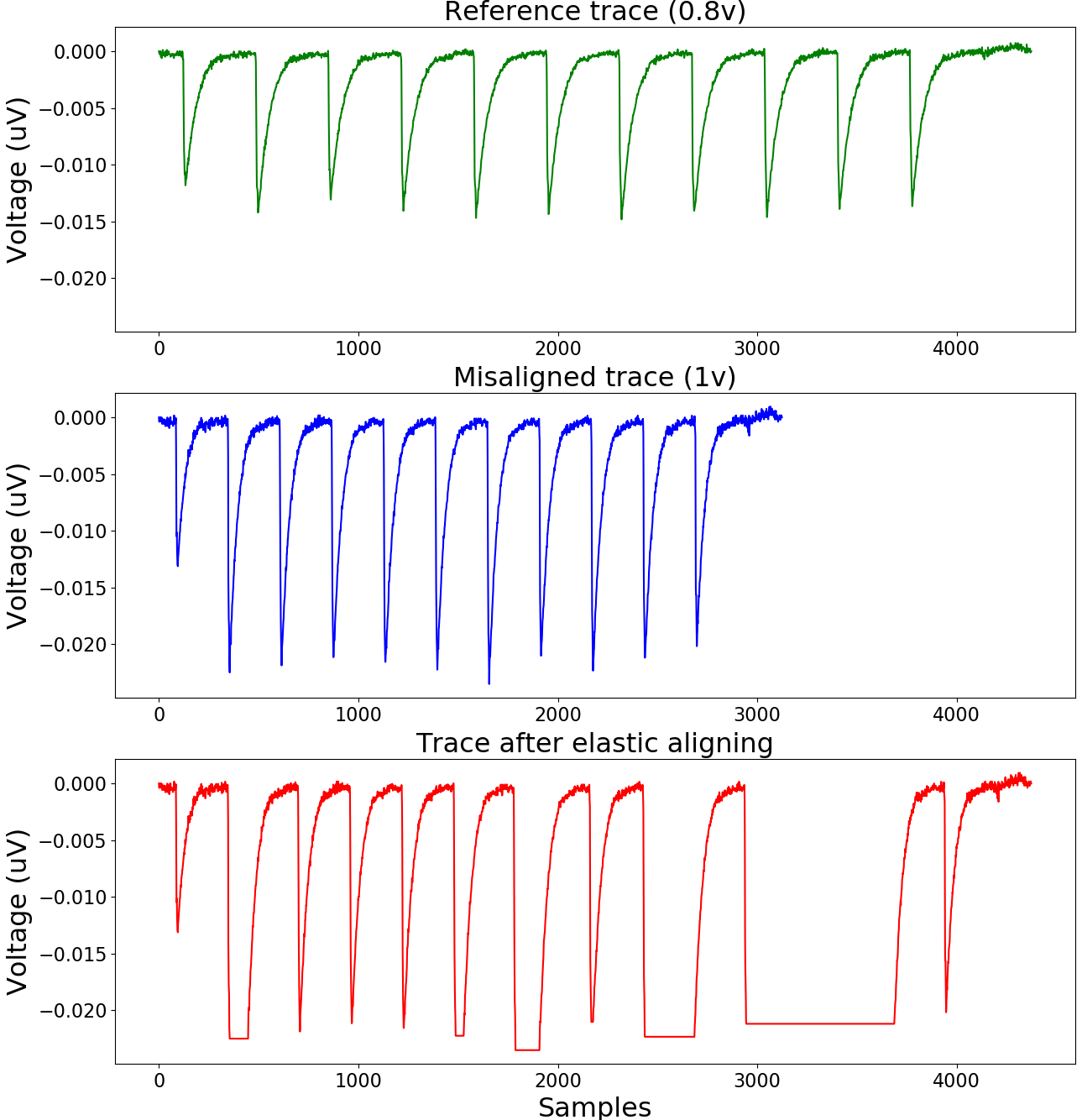}
    \caption{Example of elastic aligned trace after single-island DVS}
    %\vspace{-2mm}
    \label{fig:warping_single_dvs}
\end{figure}
However, even for the single-island DVS case, the elastic alignment does not perform nearly as well as for the frequency scaling case shown in \Cref{sec:backg}. 
As illustrated in \Cref{fig:warping_single_dvs}, the aligned trace does not match up with the reference trace. The frequency differs from that of the reference trace, and there are gaps of no activity where none actually exists in the reference.
%process
%resulting 
%figure above is traces generated from one single DVS case. Same as figure \Cref{fig:warping_single_dvs}, the top trace is the reference one, second trace is the trace for aligning and third trace is the aligning result. The result shows the aligning are not successful.
%

When we perform the same experiment with two independent voltages, the elastic technique is even less successful. The trace after alignment is almost the same as the original misaligned one, i.e., % and the warp path is close to diagonal, i.e., 
it yields negligible alignment that does not significantly help the attack.
\begin{comment}
%. The LHS of figure \Cref{fig:warping_two_islands} shows the aligning result for two islands iRDVS. The RHS of this figure presents the corresponding warp path. According to these two figures, in the two islands case, the path is close to the diagonal and this technique almost did nothing for aligning because of unhelpful warp path.
\begin{figure}[hbt!]
    \centering
    \includegraphics[width=9.5cm]{warping_two_island.jpg}
    \caption{Elastic aligning for two-islands iRDVS}
    \label{fig:warping_two_islands}
\end{figure}
\end{comment}
To further demonstrate the ineffectiveness of the elastic technique against iRDVS, we simplified iRDVS to two islands, fixed the voltage of the island to be attacked, and randomized the voltage of the other island, providing only two candidates for this random voltage. We first applied the elastic technique to these traces and then performed CPA on the elastic aligned traces.
The results
%, illustrated in \Cref{table:warping_results}, 
show that MTD remains larger than $100k$ traces and that the PGE increases from 49 to 148, suggesting that the elastic technique actually hurts the attack. 
\Cref{fig:warping_single_dvs} shows how the elastic technique fails to align operations of interest even for $n=m=1$, so for this $n=m=2$ case we would expect that these operations would be further misaligned from one another, decreasing the attack success as we observed in \Cref{fig:MTD_PGE_Combination}.

\begin{comment}

%table
\begin{table}[h]
%\captionsetup{font=normalsize}
\caption{Effectiveness of Elastic Alignment Plus CPA on Two-island iRDVS Traces}
\centering
\scalebox{1}{
\begin{tabular}{|c|c|} 
 \hline
MTD for original iRDVS traces & $>$100k  \\
\hline
Avg. PGE for original iRDVS traces & 49 \\
\hline
MTD for traces aligned by elastic technique & $>$100k \\
\hline
Avg. PGE for traces aligned by elastic technique & 148\\
\hline
\end{tabular}}

\label{table:warping_results}
\end{table}
\end{comment}
\vspace{-2mm}

\subsection{Resistance to Clustering}

To analyze the potential benefits of clustering as a preprocessing step, we varied the number of clusters $K$ and plotted MTD and PGE as functions of $K$. \Cref{fig:MTD_PGE_Combination} presents the MTD and PGE for $g=5$ with the number of independent islands $n$ ranging from one to four. $100k$ traces were used for $n=1$ and $2$, and $200k$ traces were used for $n=3$ and $4$, to account for the decreased cluster size described in \Cref{sec:clusteringattack}. We assume that $m=n$ for every experiment.

For the single-island DVS case, using clustering reduces the MTD from over $100k$ ($K=1)$ down to $16k$. This minimum is achieved when the number of clusters is well chosen ($K=5$).
For the two-island case, when the number of clusters is close to the ideal $K=15$, the clustering attack can disclose most subkeys with $100k$ traces.
Note that as illustrated in \Cref{fig:MTD_PGE_Combination}, if the attackers are not able to correctly estimate $K$, the MTD exceeds $100k$.
For three-island iRDVS, for all values tested, the MTD was between $100k$ and $200k$. The PGE reaches 0 when the number of clusters reaches the ideal $K=35$, showing that the secret is disclosed when $K$ is correctly specified. %nearing disclosure and would be uncovered with more traces.
For the four-island case $K$ was swept from 10 to 100, with the ideal being 70. The PGE at $K = 70$ is 123 and the MTD exceeds $200k$, indicating that the secret is far from being uncovered.

\section{Measurements of an AES chip}
\label{sec:chip}
\begin{figure}[hbt!]
    \centering
    \includegraphics[width=0.6\columnwidth]{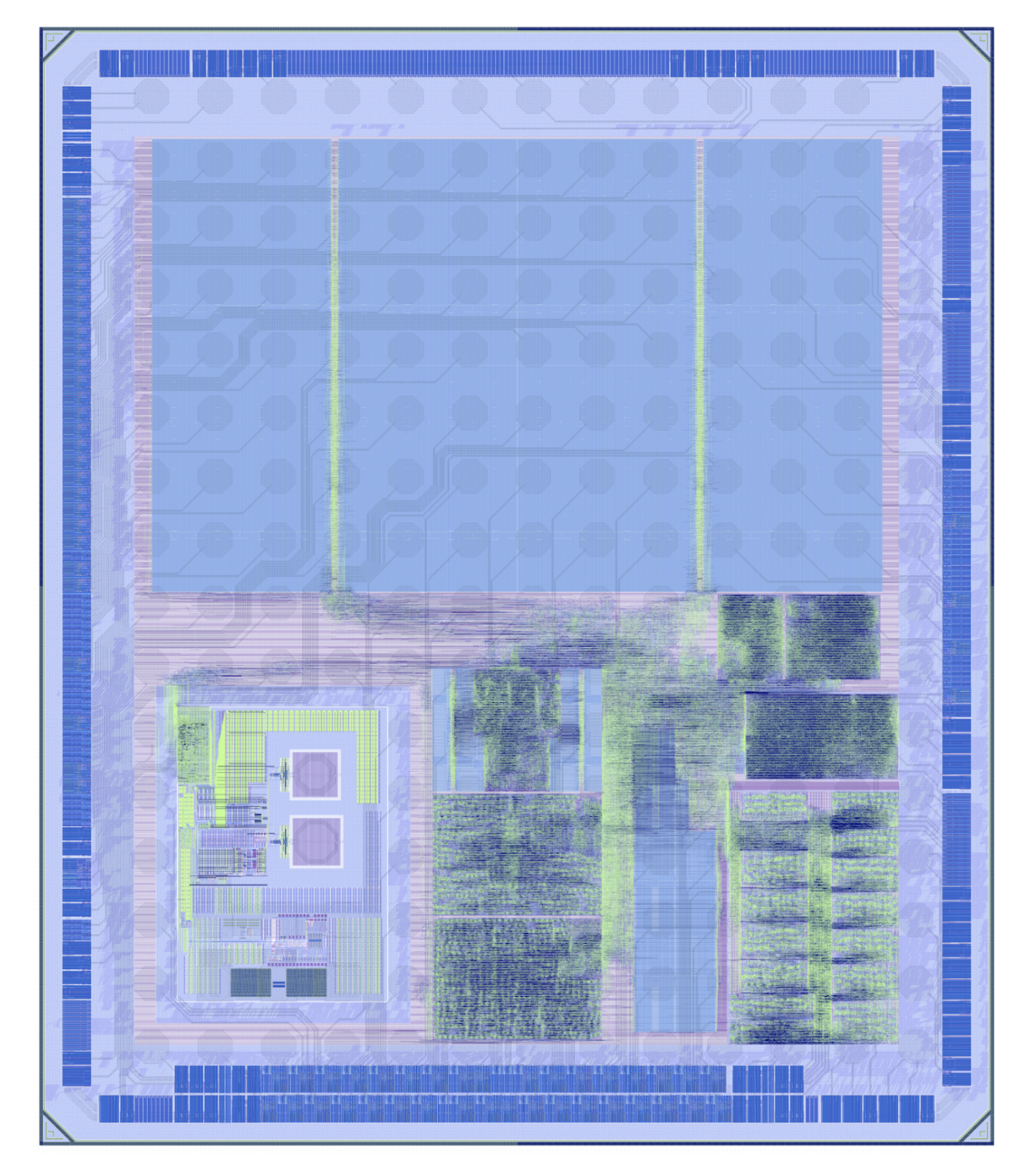}
    \caption{Die plot of the manufactured iRDVS chip}
    \label{fig:die_graph}
\end{figure}
A chip was manufactured in a 12nm FinFet process to demonstrate the performance of the approach in silicon, Figure~\ref{fig:die_graph} presents its die plot. The chip included a 1GHz RISCV complex for control and several variants of the AES core, both synchronous and asynchronous, including one with iRDVS protection. We found that with the AES cores running at 1GHz, it requires significant effort to identify precisely at what time the encryption batch starts and ends, and further effort to separate the encryption batch into individual encryptions. Previous analyses on simulated data expected separated encryption traces, so we modified the TVLA test to be able to perform the analysis without needing to separate encryption traces. The TVLA test was chosen because it does not require a successful or near-successful attack to obtain security results, and the non-specific test is attack-agnostic. Running one encryption at a time would not show the benefits of iRDVS, because iRDVS relies on several islands operating at once for obfuscation, which requires several encryptions to be in different stages of the AES pipeline at once. While the traditional non-specific fixed-vs-random TVLA test compares the encryption of one fixed plaintext with the encryption of one “random” plaintext, we changed that one encryption to 32 encryptions, with the plaintext of interest (either fixed or random) in the middle of those 32 encryptions and the 31 other plaintexts being random. This ensures that the pipeline is as full as possible. The TVLA analysis was performed on the entire duration of those 32 encryptions. Each trace thus comprises a few thousand samples, with the number of samples varying proportionally with the time the core takes to complete the 32 operations (typically between 5K and 15K samples). 
%The outputs of the TVLA non-specific test are t-scores, which are a measure of how different the fixed and random encryption are from one another. The more different the encryptions are, the more we would expect the device to exhibit leakage that would make a power analysis attack more likely to succeed. A higher t-score indicates that the difference between the fixed and random traces is less likely to have occurred by chance. T-scores are calculated point-wise, so each sample has its own t-score. 
To account for traces being different lengths due to natural variation and due to different voltage supply levels affecting the speed of encryption, the traces were linearly interpolated to all be of the same length. Traces are split into two groups of the same number of encryptions, and the t-test is run on both groups, each comprising half of the fixed and half of the random traces. The test then compares the fixed and random traces in each group. The TVLA methodology recommends a confidence value C of 4.5, which means that any samples with t-scores < -4.5 or > 4.5 in both groups indicate that the device is insecure and “fails'' with a confidence of 99.99\%. To further demonstrate the strength of the iRDVS approach, the same analysis was performed for a C value of 2 and the results are summarized in Table ~\ref{table:tvla_results}.

In particular, our TVLA analysis compared four cases: constant voltage, DVS, adjacent iRDVS, and alternating iRDVS, all running on the iRDVS AES core. The constant voltage case runs all encryptions at a constant of 0.8V. The DVS case keeps all islands at the same voltage but changes that voltage randomly to a value between 0.6V and 0.8V for every encryption batch. The adjacent and alternating iRDVS configurations use the full capability of the iRDVS core and change the voltage of each of the four power domains randomly and independently for every encryption batch. These adjacent and alternating cases extend the same cases from the simulation experiments to use four independent voltage domains rather than two. We predict that stages with similar delay maximize overlapping computation and minimize potential timing side channels. For this reason, every other flip-flop is transparent, creating seven pipeline stages, each with two rounds of AES and two voltage islands.

\vspace{1mm}
%table
\begin{table}[h]
%\captionsetup{font=normalsize}
\caption{Non-specific fixed-vs.-random TVLA results - \# of samples with t-scores exceeding C}
\centering
\scalebox{1}{
\begin{tabular}{|c|c|c|} 
 \hline
Case & C = 4.5 & C = 2.0 \\
\hline
Constant & 740  & 2211 \\
\hline
DVS & 0 & 34\\
\hline
Adjacent iRDVS & 0 & 12\\
\hline
Alternating iRDVS & 0 & 4\\
\hline
\end{tabular}}

\label{table:tvla_results}
\end{table}
\vspace{-5mm}

\section{Conclusions and Future Work}
\label{sec:concl}
% we may estimate overhead in conclusion

According to mathematical analysis and experimental results, the proposed island-based random DVS approach not only maintains the merits of DVS but also introduces misalignment that is resistant to advanced alignment techniques. The proposed unsupervised machine learning based attack is successful with three or fewer islands but increases in difficulty as the number of islands grows. This is in part because the number of voltage supply settings grows combinatorially with the number of independent voltages. 

\begin{comment}
Importantly, the synthesized multi-island DVS traces analyzed here include regions where some islands remain active longer than others due to their lower voltages, which makes them more vulnerable. This would not be true in the proposed design introduced in \Cref{sec:irdvs}, where half of the AES rounds are operating in parallel and all pipeline stages have the same delay. We would thus expect this proposed design to be even more resistant to clustering. 
\end{comment}

An AES iRDVS chip was fabricated to demonstrate the security benefits of the approach, using the industry-accepted TVLA methodology and resulting metric. Further work can be done with larger sample sizes, and work can be done to separate encryption traces precisely so that individual encryptions can be analyzed with TVLA and attacked with CPA or other power analysis attacks. The DVS and adjacent and alternating island-based RDVS cases comprise only a small portion of the potential design space for the iRDVS AES core. Our future work includes finding the most secure configuration as well as finding the configurations that best balances security, performance, and power. Future efforts will also include the addition of voltage generation and randomization circuitry as part of the core. This will further improve the security of the system by reducing the attack surface of the core.

\section{Acknowlegement}
This work, including the fabrication of the chip presented, was supported by DARPA under the "21 Century Cryptography" program and contract \#HR001119C0070. 
%We are currently working on the design of an AES accelerator that uses this technique, which will have a configurable number of independent voltage islands; its tapeout is finished and testing is pending.
%Using this accelerator, we plan to evaluate the effectiveness of the iRDVS technique against CPA using real traces.  
%Additional future work includes addressing the resistance to other forms of attack, including EM attacks that may try to discover and track the voltage of each island. In particular, we are exploring novel physical design strategies that will thwart such efforts.
\bibliographystyle{ACM-Reference-Format}
\bibliography{reference}

\end{document}